\documentclass[12pt,tightenlines,preprintnumbers,%
superscriptaddress,nofootinbib]{revtex4}
\newcommand{\PRE}[1]{{#1}} % Use if preprint style
\usepackage{bm} \usepackage{epsfig}
\newcommand{\eqref}[1]{Eq.~(\ref{#1})}

\begin{document}

\preprint{preprint MIT-CTP 3709}  

\title{

Enlightenment, Knowledge, Ignorance, Temptation\footnote{Summary talk at ``Expectations of a Final Theory'', Trinity College, Cambridge, September 2005.  Drastically condensed and otherwise modified from the talk as delivered.}

}

\PRE{\vspace*{.5in}}

\author{Frank Wilczek%
\PRE{\vspace*{.2in}}
} 
\affiliation{Center for Theoretical Physics, Department of Physics,
Massachusetts Institute of Technology, Cambridge, Massachusetts 02139,
USA
\PRE{\vspace*{.5in}}
}

%\date{May 2005}

%\pacs{12.60.Jv, 04.65.+e, 95.35.+d, 13.85.-t}
%12.60.Jv Supersymmetric models
%04.65.+e Supergravity
%95.35.+d Dark matter 
%13.85.-t Hadron-induced high- and super-high-energy interactions 
%             (energy>10 GeV)

\begin{abstract}
I discuss the historical and conceptual roots of reasoning about the parameters of fundamental physics and cosmology based on selection effects.   I argue concretely that such reasoning can and should be combined with arguments based on symmetry and dynamics; it supplements them, but does not replace them.  
\end{abstract}

\maketitle

\PRE{\newpage}

\section{A New Zeitgeist}

Our previous Rees-fest ``Anthropic Arguments in Fundamental Physics and Cosmology'' at Cambridge in 2001 had much in common with this one, in terms of the problems discussed and the approach to them.  Then as now the central concerns were apparent conspiracies among fundamental parameters of physics and cosmology that appear necessary to insure the emergence of life.  Then as now the main approach was to consider the possibility that significant observational selection effects are at work, even for the determination of superficially fundamental, universal parameters.    

That approach is loosely referred to as anthropic reasoning, which in turn is often loosely phrased as the anthropic principle: the parameters of physics and cosmology have the values they do in order that intelligent life capable of observing those values can emerge.   That formulation upsets many scientists, and rightly so, since it smacks of irrational mysticism.    

On the other hand, it is simply a true fact that intelligent observers are located only in a miniscule fraction of the world, and in places with special properties.  As a trivial consequence, probabilities conditioned on the presence of observers will differ grossly from probabilities per unit volume.   Much finer distinctions are possible, and useful; but I trust that this word to the wise is enough to it make clear that we shouldn't turn away from straightforward logic just because it can be made to sound, when stated sloppily, like irrational mysticism.   

For all their commonality of content, the spirit pervading the two gatherings seemed quite different, at least to me.   One sign of the change is the different name attached to the present gathering.  This time it's ``Expectations of a Final Theory''.    

The previous gathering had a defensive air.  It prominently featured a number of physicists who subsisted on the fringes, voices in the wilderness who had for many years promoted strange arguments about conspiracies among fundamental constants and alternative universes.   Their concerns and approaches seemed totally alien to the consensus vanguard of theoretical physics, which was busy successfully ;-) constructing a unique and mathematically perfect Universe.   

Now the vanguard has marched off to join the prophets in the wilderness.    According to the new zeitgeist, the real world of phenomena must be consulted after all, if only to position ourselves within a perfect, but inaccessible, Multiverse.   Estimating selection effects, in practice, requires considerations of quite a different character than what we've become accustomed to in the recent practice of theoretical (i.e., hep-th) physics:  looser and more phenomenological, less precise but more accurate.

\section{Sources}

What caused the change?   

In his opening talk \cite{weinberg} Steve Weinberg ascribed the change in attitude to recent developments in string theory, but I think its deep roots mostly lie elsewhere and go much further back in time.   Those of us who attended ``Anthropic Arguments''  lived through an empirical proof of that point.   I'd like to elaborate on this issue a little, not only as a matter of accurate intellectual history, but also to emphasize that the main arguments do not rely on narrow, delicate (I'd venture to say fragile) technical developments: rather, they are broadly based and robust.

\begin{description}
\item[1. The standardization of models:] 
With the extraordinary success of the standard model of fundamental physics, brought to a new level of precision at LEP through the 1990s; and with the emergence of a standard model of cosmology, confirmed by precision measurements of microwave background anisotropies, it became clear that an excellent working description of the world as we find it is in place.  In particular, the foundational laws of physics that are relevant to chemistry and biology seem pretty clearly to be in place.  

The standard models are founded upon broad principles of symmetry and dynamics, assuming the values of a handful of numerical parameters as inputs.   Given this framework, we can consider in quite an orderly way the effect of a broad class of plausible changes in the structure of the world: namely, change the numerical values of those parameters!   When we try this we find, in several different cases, that the emergence of complex structures capable of supporting intelligent observation appears quite fragile.  

On the other hand, valiant attempts to derive the values of the relevant parameters, using symmetry principles and dynamics, have not enjoyed much success.   

Thus life appears to depend upon delicate coincidences that we haven't been able to explain.  The broad outlines of that situation have been apparent for many decades.  When less was known it seemed reasonable to hope that better understanding of symmetry and dynamics would clear things up.  Now that hope seems much less reasonable.  The happy coincidences between life's requirements and nature's choices of parameter-values might be just a series of flukes, but one could be forgiven for beginning to suspect that something deeper is at work.   

That suspicion is the first deep root of anthropic reasoning. 

\item[2. The exaltation of inflation:]

The most profound result of observational cosmology has been to establish the Cosmological Principle: that the same laws apply to all parts of the observed universe, and moreover matter is, on average, uniformly distributed throughout.   

It seems only reasonable, then, to think that the observed laws are indeed Universal, allowing no meaningful alternative, and to seek a unique explanation for each and every aspect of them.   Within that framework, explanations invoking selection effects are moot.  If there is no variation, then there cannot be selection. 

Inflationary cosmology challenges that interpretation.  It proposes a different explanation of the Cosmological Principle: that the observed universe originated from a small patch, and had its inhomogeneities ironed out dynamically.   In most theoretical embodiments of inflationary cosmology, the currently observed universe appears as a small part of a much larger Multiverse.   In this framework observed universal laws need not be Universal (that is, Multiversal), and it is valid -- indeed, necessary --  to consider selection effects. 

The success of inflationary cosmology is the second deep root of anthropic reasoning.

\item[3. The unbearable lightness of space-time:]

Among the coincidences between life's requirements and nature's choices of parameter-values, the smallness of the cosmological term, relative to its natural value, is especially clear and striking.   

Modern theories of fundamental physics posit an enormous amount of structure within what we perceive as empty space: quantum fluctuations, quark-antiquark condensates, Higgs fields, and more.   At least within the framework of general relativity, gravity responds to every sort of energy-momentum, and simple dimensional estimates of the contributions from these different sources suggest values of the vacuum energy, or cosmological term, many orders of magnitude larger than what is observed.   Depending on your assumptions, the discrepancy might involve a factor of $10^{60}, 10^{120}$, or $\infty$.  

Again, attempts to derive an unexpectedly small value for this parameter, the vacuum energy, have not met with success.   Indeed most of those attempts aimed to derive the value zero, which now appears to be the wrong answer.   

In 1987 Weinberg proposed to cut this Gordian knot by applying anthropic reasoning to the cosmological term.  On this basis he predicted that the cosmological term, rather than being zero, would be as large as it could be, while remaining consistent with the emergence of observers.     The numerical accuracy of this prediction is not overwhelmingly impressive (the computed probability to observe a cosmological term as small as we do is roughly 10\%), though this might be laid to the vagaries of sampling a statistical distribution just once.   Also the original calculation was based on the hypothesis that one should consider variations in the vacuum energy alone, keeping all other parameters fixed, which might be too drastic a simplification.   In any case, the apparent observation of vacuum energy that is ridiculously small from a microphysical perspective, but importantly large from a cosmological perspective, certainly encourages explanation based on selection.

\item[4. The superabundance of string theory:]

After a brief, heady period around 1984-5, during which it seemed that simple general requirements (e.g., $N=1$ supersymmetry and three light fermion generations) might pick out a unique Calabi-Yau compactification as the description of observed reality, serious phenomenological application of string theory has been forestalled by the appearance of a plethora of candidate solutions.    The solutions all exhibited unrealistic features (e.g. unbroken supersymmetry, extraneous massless moduli fields), and it was anticipated that when those problems were fixed some degree of uniqueness might be restored.    It was also hoped that string theory would provide a dynamical understanding for why the cosmological term is zero.  

Recent constructions have provided a plethora of approximate solutions with broken supersymmetry and few or no moduli fields.   They are not stable, but it is plausible that some of them are metastable with very long lifetimes indeed.   As yet none (among $\gtrsim 10^{\rm hundreds}$) appears to be entirely realistic, but there's still plenty of scope for investigation in that direction, and even for additional constructions.    

In these new constructions the cosmological term can take a wide range of values, positive or negative.   So if cosmology provides a Multiverse in which a significant sample of these metastable solutions are realized, then the stage might be set for selection effects to explain (roughly) the value we actually observe, as I just sketched.

\end{description}

\section{Losses}

Einstein expressed the traditional, maximally ambitious vision of mathematical physics with characteristic lucidity
\begin{quote}
I would like to state a theorem which at present can not be based upon anything more than upon a faith in the simplicity, i.e., intelligibility, of nature: there are no arbitrary constants ... that is to say, nature is so constituted that it is possible logically to lay down such strongly determined laws that within these laws only rationally completely determined constants occur (not constants, therefore, whose numerical value could be changed without destroying the theory).
\end{quote}

Over the course of the twentieth century, that program has worked remarkably well.  Rather than waste words to belabor the point, I'll just present you with three icons %(Figures 1-3).   
\begin{center}
\includegraphics[width=3in]{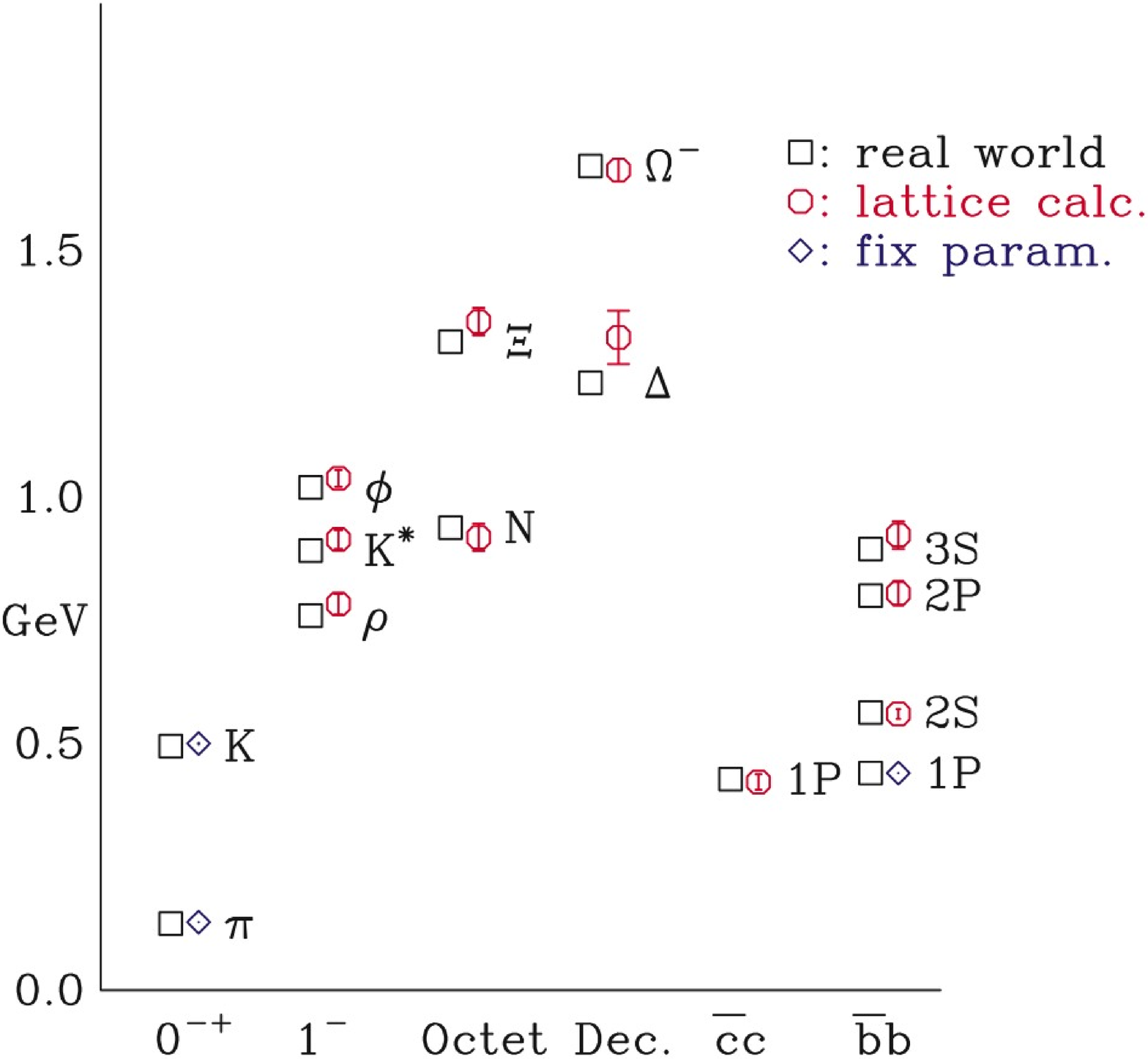}\\
{\small Figure 1.  The calculation of particle masses in QCD.  ``N'' denotes nucleon.  These calculations, which employ the full power of modern computers, account for the bulk of the mass of ordinary matter on the basis of a conceptually based yet fully algorithmic theory.}
\label{fig1}\\[1ex]

\includegraphics[width=3in]{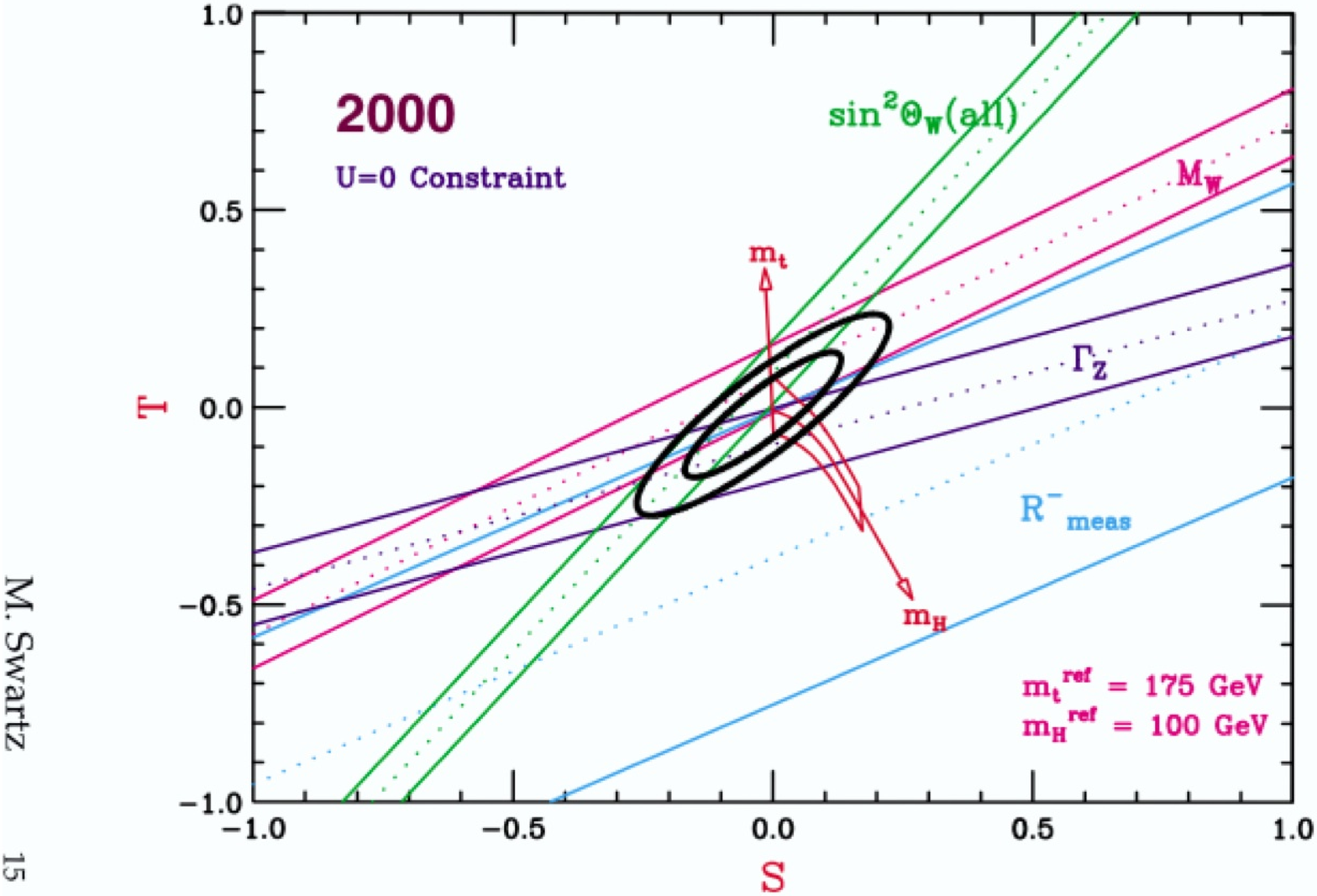}
\label{fig2}\\
{\small Figure 2.  Overdetermined, precision comparison of theory and experiment in electroweak theory, including radiative corrections.  The calculations make ample use of the intricate rules for dealing with virtual particles in quantum field theory.  Successful confrontations between theory with experiment, of the sort shown here and in Figure 1, ground our standard model of fundamental interactions.}\\[1ex]

\includegraphics[width=3in]{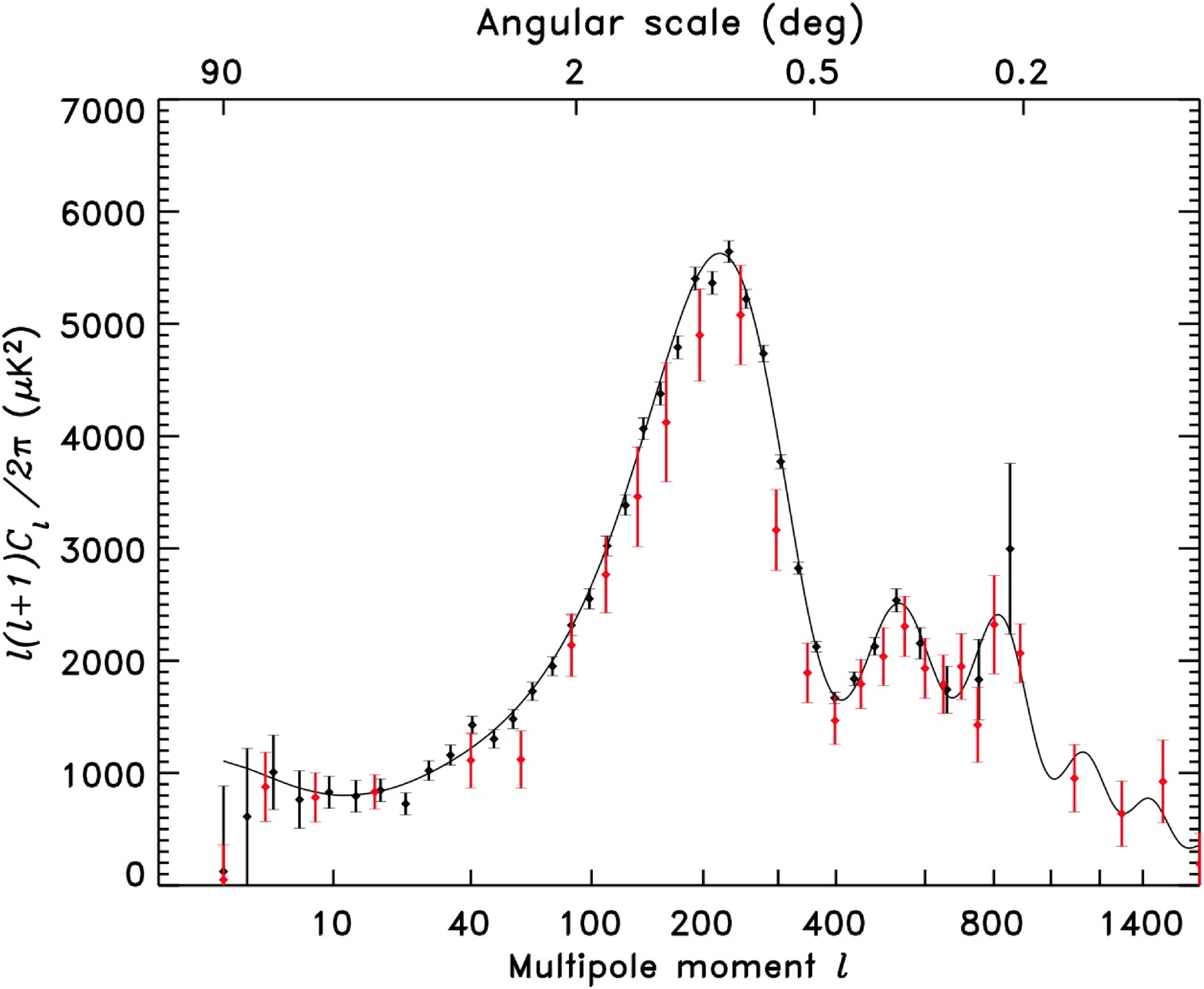}\label{fig3}\\
{\small Figure 3.  Comparison of standard cosmological model including dark matter, dark energy, and scale invariant, adiabatic, Gaussian fluctuation spectrum with observed microwave anisotropies.   The successful confrontation of theory and experiment in this case ground our new standard model of cosmology.  It traces the origin of all macroscopic structure to growth of simply characterized, tiny seed fluctuations through gravitational instability.}
\end{center}

What's most characteristic of these icons is their richness of detail and their quantitative precision.  They confront profound theoretical ideas and complex calculations with concrete, precise observations.   The fact that we physicists can worry over possible discrepancies at the level of parts per billion, in the case of the muon's magnetic moment, is our unique glory.    Such examples epitomize what, traditionally, has distinguished fundamental physics from softer, ``environmental''  disciplines like history and biology. 

With those words and images in mind, let me lament our prospective losses, if we adopt anthropic or statistical selection arguments too freely:

\begin{description}
\item[1. Loss of precision:]
I don't see any realistic prospect that anthropic or statistical selection arguments -- applied to a single sampling! -- will ever lead to anything  comparable in intellectual depth and numerical precision to what these icons represent.  In that sense, intrusion of selection arguments into foundational physics and cosmology really does, to me, represent a genuine lowering of expectations.
\item[2. Loss of targets:]
Because the standard models of fundamental physics and cosmology describe the world so well, a major part of what ideas going beyond those standard models could aspire to achieve, for improving our understanding of the world, would be to fix the values of their remaining free parameters.  If we compromise on that aspiration, there will be much less about the physical world for fundamental theory to target.   
\end{description}

\section{A Classification}

Of course, physicists have had to adjust their expectations before.   In the development of Copernican-Newtonian celestial mechanics attractive {\it a priori\/} ideas about the perfect shape of planetary orbits (Ptolemy) and their origin in pure geometry (Kepler) had to be sacrificed.  In the development of quantum mechanics, ideas of strict determinism (Einstein) had to be sacrificed.   In those cases, sacrifice of appealing philosophical ideas was compensated by the emergence of powerful theories that described many specific features of the natural world and made surprising, impressive predictions.   In America we have the saying ``No pain, no gain.''   

There's a big difference, however, between those episodes and the present one.  
Resort to anthropic reasoning involves plenty of pain, as I've lamented, but so far the gain has been relatively meagre, to say the least.   

Even if we can't be precise in our predictions of fundamental parameters, we can still aspire to clear thinking.  Specifically, we can try to be clear concerning what it is we can or can't be precise about.   In this way we can limit our losses, or at least sharpen our discussion.  In that spirit, I'd like to suggest a chart (Figures 4 and 5) that draws some helpful boundaries.
%\begin{minipage}[here]{6.5in}
%\includegraphics[width=3in]{frankimage4.eps}\label{fig:4}\quad\ 
%\includegraphics[width=3in]{newfrankimage5.eps}\label{default}\\
%{\small Fig. 3}
%\end{minipage}

\begin{center}
\includegraphics[width=3in]{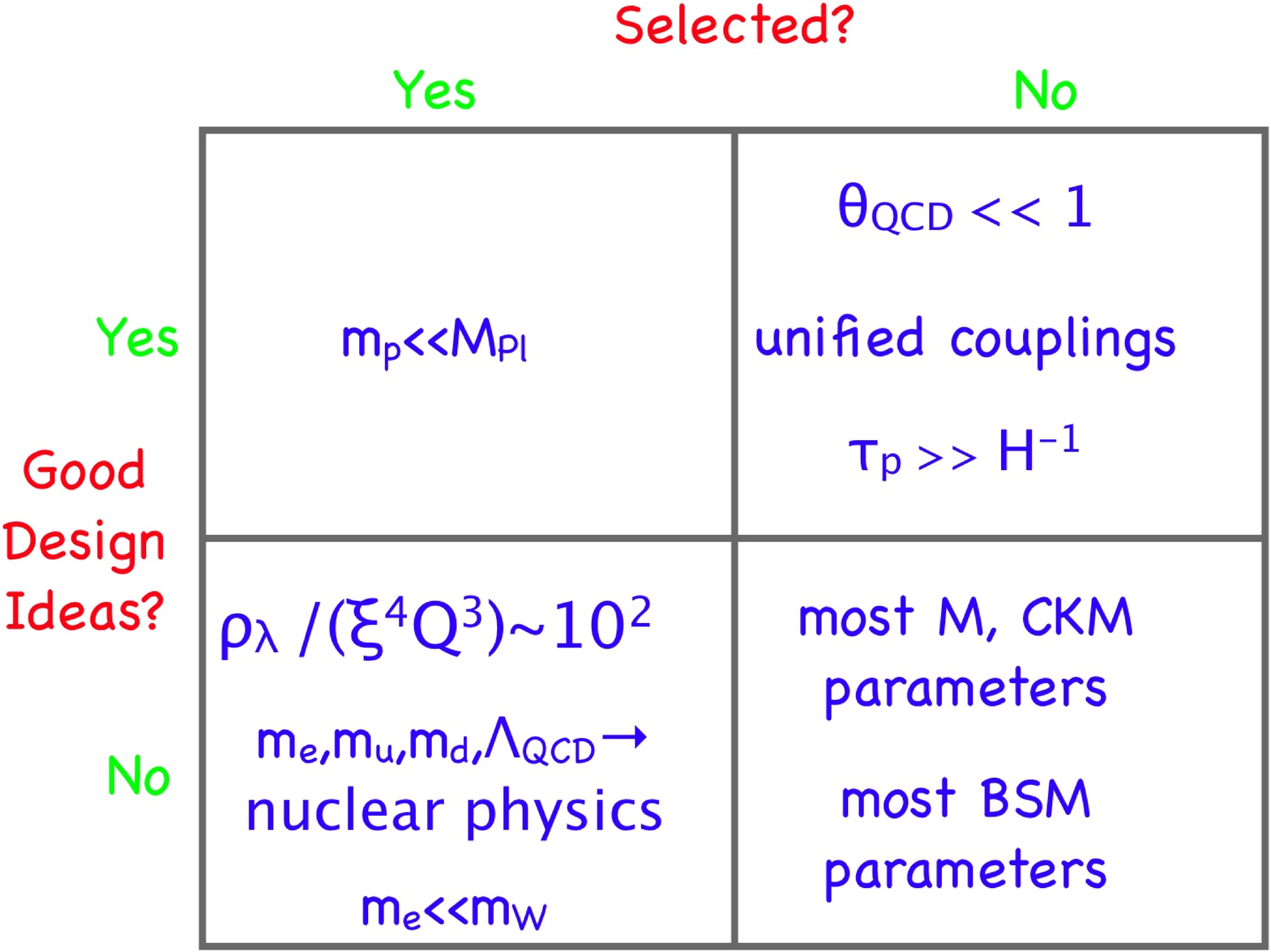}\\
{\small Figure 4.  Classification of various parameters by criteria of selective pressure and theoretical insight.  For definitions and discussion, see the text.}
\label{fig4}\\[1ex]

\includegraphics[width=3in]{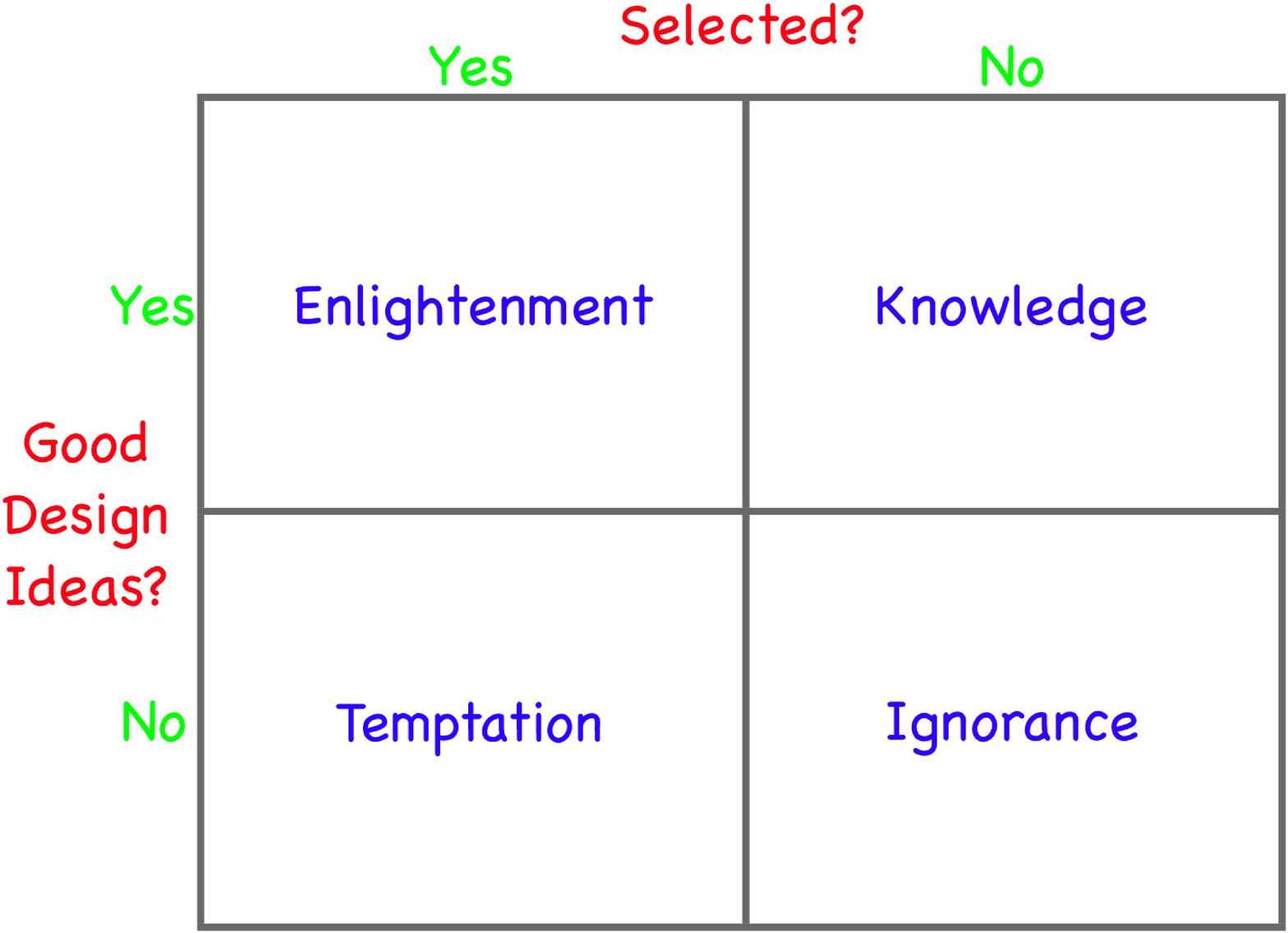}\\
{\small Figure 5. Appropriate names for the different classes of parameters.  ``Temptation'' labels the natural habitat for anthropic reasoning.}
\label{fig5}\\[1ex]
\end{center}

The chart  provides four boxes wherein to house parameters, or salient combinations of parameters, as in Figure 4.  On the horizontal axis, we have a binary distinction: is the parameter selected for, in the sense of anthropic reasoning, or not?  In other words, is it relevant to the emergence of intelligent life, or not?  On the vertical axis, we have a different binary distinction: is the parameter one about whose values we have promising ideas based on symmetry and dynamics, or not?   In that way we divide up the parameters into four classes.  In Figure 5, I've named the different classes.  
\begin{description}
\item[1. Enlightenment:] This class contains salient combinations of parameters that are both crucial to life, and at least {\it significantly\/} understood.   Its box is rather sparsely populated.  I've entered the tiny ratio of the proton mass m$_{\rm p}$ to the Planck mass M$_{\rm Pl}$.  That small ratio is what allows the pull and tug of nuclear physics and chemistry, with attendant complexity, to dominate the relentless crunch of gravity.   It can be understood as a consequence of the logarithmic running of the strong coupling, and the $SU(2)\times U(1)$ veto of quark and electron masses, modulo the weak-scale hierarchy problem (which opens a can of worms).  

\item[2. Knowledge:] This class contains parameters or regularities that do not appear to be crucial for life, but have been interpreted to have profound theoretical significance.   Among these are the tiny $\theta$ parameter of QCD, the relationship among low-energy $SU(3)$, $SU(2)$, and $U(1)$ couplings that enables their unification at high energy, and the extremely long lifetime of the proton.  

The $\theta$ parameter encodes the possibility that QCD might support violation of parity (P) and time reversal (T) in the strong interaction.   It is a pure number, defined modulo $2\pi$. Experimental constraints on this parameter require $|\theta| \lesssim 10^{-9}$, but it is difficult to imagine that life requires better than $|\theta | \lesssim 10^{-1}$, if that, since the practical consequences of nuclear P and T violation seem insignificant.  On the other hand there is a nice theoretical idea, Peccei-Quinn symmetry, that could explain the smallness of $\theta$.   Peccei-Quinn symmetry requires expansion of the standard model, and implies the existence of a remarkable new particle, the axion, of which more below.    

The unification relationship among gauge couplings encourages us to think that the corresponding gauge symmetries are aspects of a single encompassing symmetry which is spontaneously broken, but would become manifest at asymptotically large energies, or short distances.    Accurate quantitative realization of this idea requires expanding the standard model even at low energies.  Low-energy supersymmetry in any of its forms (including focus point or split supersymmetry) is very helpful in this regard.   Low-energy supersymmetry of course requires a host of new particles, some of which should materialize at the LHC. 

It is difficult to see how having the proton lifetime $\tau_p \gg 10^{18}$ seconds, the lifetime of the universe, could be important to life.  Yet the observed lifetime is at least $\tau_p \gtrsim 10^{40}$ seconds.   Conventional anthropic reasoning is inadequate to explain that observation.  (Perhaps if we included potential {\it future\/} observers in the weighting??)   On the other hand the unification of couplings calculation implies a very high energy scale for unification, which supplies a natural suppression mechanism.   Detailed model implementations, however, suggest that if gauge unification ideas are on the right track, proton decay should occur at rates not far below existing limits.   
 
\item[3. Ignorance:] This class contains parameters that are neither important to life, nor close to being understood theoretically.  It includes the masses $M$ and weak mixing angles  of the heavier quarks and leptons (encoded in the Cabibbo-Kobayashi-Maskawa, or CKM, matrix), and the masses and mixing angles of neutrinos.  It also includes most of the prospective parameters of models beyond the standard model (BSM), such as low-energy supersymmetry, because only a few specific properties of those models (e.g., the rate of baryogenesis) are relevant to life.  Of course, if the Multiverse supports  enough  variation to allow selection to operate among a significant fraction of the parameters that are relevant to life, there is every reason to expect  variation also among some parameters that are not relevant to life.   In an abundant Multiverse, wherein any particular location requires specification of many independent coordiantes, we might expect this box to be densely populated, as evidently it is. 

\item[4. Temptation:] This class contains parameters whose values are important to life, and are therefore subject to selection effects, but which look finely tuned or otherwise odd from the point of view of symmetry and dynamics.    It is in understanding these parameters that we are tempted to invoke anthropic reasoning.   This class includes the smallness of the dark energy, mentioned previously, and several other items indicated in Figure 5.   

Life in anything close to the form we know it requires both that there should be a complex spectrum of stable nuclei, and that the nuclei can get synthesized in stars.  As emphasized by Hogan \cite{hogan}  and many others,  those requirements imply constraints, some quite stringent, relating the QCD parameters $\Lambda_{\rm QCD}, m_u, m_d$ and $m_e$ and $\alpha$.   On the other hand these parameters appear on very different footings within the standard model and in existing concrete ideas about extending the standard model.   The required conspiracies {\it among\/} the masses $m_u, m_d, m_e$ are all 
the more perplexing because each of the masses is far smaller than the ``natural'' value, ~250 GeV, set by the Higgs condensate.  An objective measure of how unnatural this is, is that pure-number Yukawa couplings of order $10^{-6}$ underlie these masses.   

More recent is the realization that the emergence of user-friendly macrostructures, that is stable planetary systems, requires rather special relationships among the parameters of the cosmological standard model.   Here again, no conventional symmetry or dynamical mechanism has been proposed to explain those relationships; indeed, they connect parameters whose status within existing microscropic models is wildly different.  Considerations of this sort have a rich literature, beginning with \cite{carr}.  
Reference \cite{axion2} contains detailed discussion of these matters, which bring in some very interesting astrophysics.  (In this regard \cite{axion2} greatly improves on \cite{axion1}, and on my summary talk as actually delivered.)   A major result of that paper is a possible anthropic explanation of the observed abundance $\xi$ of dark matter, conditioned on the density of dark energy $\rho_{\lambda}$ and the amplitude $Q$ of primeval density fluctuations. 
\end{description}

Dynamical versus anthropic reasoning is not an either/or proposition.   It may be that some parameters are best understood dynamically and others anthropically (and others not at all).   In my chart, no box is empty.   

Indeed, there is much potential for fertile interaction between these different modes of reasoning.   For example both axion physics and low-energy supersymmetry provide candidates for dark matter, and dark matter has extremely important anthropic implications \cite{axion2}. 

Nor is the situation necessarily static.  We can look forward to a flow of parameters along the paths from Ignorance to Enlightenment as physics progresses.

\section{A New Zeitgeist?}

Actually, it's quite old. 

Earlier I discussed ``losses''.   There could be a compensating moral gain, however, in well-earned humility.   What we are ``losing'', we never really had.   Pure thought didn't supersede creative engagement with phenomena as a way of understanding the world twenty years ago, hasn't in the meantime, and won't anytime soon. 

I think it's been poetic to witness here at Trinity a re-emergence of some of the spirit of Newton.  Perhaps not yet
\begin{quote}
Hypothesis non fingo.
\end{quote}
but I hope
\begin{quote}
I know not how I seem to others, but to myself I am but a small child wandering upon the vast shores of knowledge, every now and then finding a small bright pebble to content myself with while the vast ocean of undiscovered truth lay before me.
\end{quote}

\section*{Acknowledgments}

The work of FW is supported in part by funds provided by
the U.S. Department of Energy under cooperative research agreement
DE-FC02-94ER40818.

%%%%%%%%%%%%%%%%%%%%%%%%%%%%%%%%%%%%%%%%%%%%%%%%%%%%%%

%%%%%%%%%%%%%%%%%%%%%%%%%%%%%%%%%%%%%%%%%%%%%%%%%%%%%%

\end{document}